\RequirePackage{fix-cm}
\documentclass[twocolumn]{svjour3}          % twocolumn
\smartqed  % flush right qed marks, e.g. at end of proof
\usepackage{graphicx}
\usepackage{subfigure}
\usepackage{url}
\newcommand\blfootnote[1]{%
  \begingroup
  \renewcommand\thefootnote{}\footnote{#1}%
  \addtocounter{footnote}{-1}%
  \endgroup
}
\begin{document}

\title{Non-uniqueness of solutions in asymptotically self-similar shock reflections}

\author{S. She-Ming Lau-Chapdelaine         \and
        Matei I. Radulescu %etc.
}

\institute{S. S.M. Lau-Chapdelaine\at
			  Universit\'e d'Ottawa University, \\
			  Department of Mechanical Engineering \\
              161 Louis Pasteur \\
              CBY A205\\
			  Ottawa, ON, Canada, K1N-6N5 \\
              \email{slauc076@uottawa.ca}
}

\date{Received: 29 June 2011 / Revised: 15 July 2013 / Accepted: 13 August 2013 / Published online: 24 September 2013}
\maketitle

\blfootnote{Originally published by Springer-Verlag Berlin Heidelberg. The final publication is available at {\tiny\url{http://www.springerlink.com/openurl.asp?genre=article&id=doi:10.1007/s00193-013-0469-0}} and  through \url{link.springer.com}}

\begin{abstract}
The present study addresses the self-similar problem of unsteady shock reflection on an inclined wedge. The start-up conditions are studied by modifying the wedge corner and allowing for a finite radius of curvature. It is found that the type of shock reflection observed far from the corner, namely regular or Mach reflection, depends intimately on the start-up condition, as the flow ``remembers" how it was started. Substantial differences were found. For example, the type of shock reflection for an incident shock Mach number $M=6.6$ and an isentropic exponent $\gamma =1.2$ changes from regular to Mach reflection between $44^\circ$ and $45^\circ$ when a straight wedge tip is used, while the transition for an initially curved wedge occurs between $57^\circ$ and $58^\circ$. 
\end{abstract}

\section{Introduction}
\label{intro}
The reflection of a steadily moving shock wave on a straight inclined wall has attracted much attention in recent decades due to its fundamental importance in unsteady gas dynamics and in all compressible media (gases, liquids, solids, nuclear matter, interstellar medium, granular media and detonation waves). Consider a shock wave as it travels at a velocity $D$, in the inertial frame of reference, through a stationary medium with an isentropic exponent $\gamma$ and speed of sound $c_0$. When the shock impinges upon a stationary wedge with an angle $\theta_w$ from the shock's normal, it will reflect from the wedge.

\begin{figure}
\begin{center}
	\subfigure[Regular reflection]{
	  \includegraphics[width=0.225\textwidth]{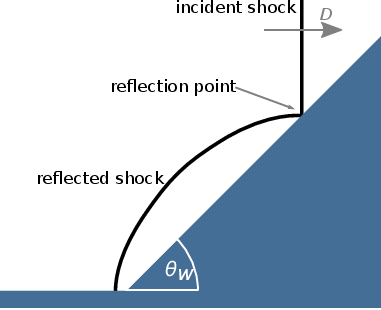}
	  \label{subfig:reflectionTypeRR}
	}
	\subfigure[Mach reflection]{
	  \includegraphics[width=0.225\textwidth]{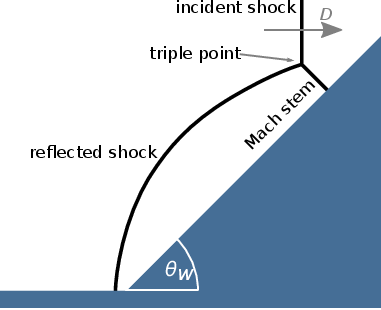}
	  \label{subfig:reflectionTypeIR}
	}
\end{center}
\caption{A moving shock, travelling at velocity $D$, strikes and reflects off of a stationary wedge of angle $\theta_w$ giving rise to a regular reflection (a) or a Mach reflection (b) depending on flow properties}
\label{fig:reflectionType}
\end{figure}

	The shock reflection will generally adopt one of two forms, either a regular or Mach reflection, exemplified by Fig. \ref{fig:reflectionType}. The Mach reflection type can be further subdivided into different regimes of reflections which have been classified by Ben-Dor \cite{ben-dor_shock_2007} and more recently by Semenov et al.  \cite{semenov_et_al_2009a,semenov_et_al_2009b} and Vasilev et al. \cite{vasilev_et_al_2008}. The isentropic exponent, the geometry (angle of the wall), and the Mach number of the incident shock $M_i=D/c_0$ uniquely define the problem; the type of reflection is subject to change with any of those variables. Thorough reviews of the shock reflection problem have been provided by Hornung \cite{hornung_regular_1986} and Ben-Dor \cite{ben-dor_shock_2007}.
	
	Experiments will typically fix the shock strength and gas type ($M_i$ and $\gamma$) and vary the wedge angle $\theta_w$. Mach reflections will occur at smaller wedge angles than regular reflections; however, there is a range of angles where both Mach and regular reflections are possible. This overlap has been observed in both steady and pseudosteady flows. Much attention has been devoted to the many types of reflection configurations, the conditions under which they occur, and which factors favour the appearance of one type over another, especially under conditions where both regular and Mach reflections are possible solutions \cite{henderson_and_menikoff_1998,henderson_and_lozzi_1975,henderson_and_lozzi_1979,hornung_et_al_1979}.

	When a shock wave reflects over a sharp corner, the absence of a characteristic length scale renders the problem self-similar.  In two-dimensions, the problem thus admits a similarity solution in terms of a combination of the space variables $x$ and $y$ and time $t$, in this case simply $x/(c_0 t)$ and $y/(c_0 t)$ \cite{jones_et_al_1951}. Indeed, such self-similar flow-fields corresponding to impulsively started gas dynamic flow-fields abound in unsteady gas dynamics (e.g., shock diffraction \cite{jones_et_al_1951}, shock refraction \cite{henderson_2001}, blast waves and implosion problems \cite{landau_and_lifshitz_1987}, etc.).

	The self-similar configuration of the reflection emerges from a singularity at the sharp tip of the reflecting surface.  When the mathematical formulation of the problem is described by the inviscid Euler equations, this does not pose any serious difficulty, as the problem is well-defined in similarity variables.  However, numerical verification of postulated self-similar solutions becomes difficult because the computational grid cannot resolve any physical or temporal singularity. Likewise, if one considers the real physical solution near the corner of finite dimension, occurring at scales commensurate with the mean free path, the problem loses its self-similarity due to the presence of physical length and time scales associated with mass, momentum, and energy transfer by molecular diffusion and the dimension of the corner. Thus no real problem is strictly self-similar.  It is only once that the evolution of the problem has evolved on length and time scales much larger than those associated with the start-up process that one may expect a weak, asymptotic form of self-similarity.

\begin{figure}
 \includegraphics[width=\columnwidth]{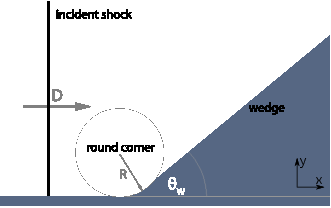}
\caption{Wedge geometry: wedge with rounded corner}
\label{fig:wedgeGeometry}
\end{figure}

	Shock reflections over obstacles with variable corner geometries (e.g., cylindrical, multiple cumulative wedges) have previously been studied \cite{henderson_and_lozzi_1979,ben-dor_and_takayama_1985,itoh_et_al_1981,ben-dor_et_al_1980,krasovskaya_and_berezkina_2008} but these studies only explored flow fields in proximity of the corner. In order to study the evolution of a shock reflection towards self-similarity, the present work focused on the problem of shock reflection over a straight wedge with a finite radius of curvature $R$ assigned to the initial corner as shown in Fig. \ref{fig:wedgeGeometry}. The evolution of the shock reflection was studied over the straight portion of the wedge over length scales much larger than $R$.  The objective was to determine whether the solution of the asymptotically self-similar problem, in the limit that $R$, relative to the distance travelled, becomes infinitesimally small, agreed with the solution of the self-similar problem, particularly in the regime of shock reflections where both regular and Mach reflections are possible.

%---------------------------------------------------------------------------------------------------------------------
\section{Problem definition and methodology}
\label{sec:method}
\subsection{Problem definition}
The problem was addressed through numerical experiments.  The reflection problem was modeled by the Euler equations for a calorically perfect gas
\begin{equation}
\frac{\partial W}{\partial t} + \frac{\partial F}{\partial x} + \frac{\partial G}{\partial y}=0
\end{equation}
with
\begin{equation}
W= \left[ \begin{array}{ccc}
	\rho \\
	\rho u \\
	\rho v \\
	E \\
\end{array}	\right] \textrm{,\ \ }
F= \left[ \begin{array}{ccc}
 	\rho u \\
	\rho u^2+p \\
	\rho uv \\
	(E+p)u \\
\end{array}	\right] \textrm{,\ \ }
G= \left[ \begin{array}{ccc}
	\rho v \\
	\rho vu \\
	\rho v^2+p \\
	(E+p)v \\
\end{array}	\right]
\end{equation}
where $\rho$ is the density,  $u$ is the $x$-component of velocity, $v$ is the $y$-component of velocity and $E$ is the total energy, given by 
\begin{equation}
E=\frac{p}{\gamma -1} +\frac{1}{2} \rho u^2+\frac{1}{2} \rho v^2 .
\end{equation}

Consistent with the absence of viscous terms in the Euler equations, the boundary conditions along the solid surfaces were reflective with vanishing normal velocity (no penetration) but allowed tangential velocities.  

The initial conditions consisted of a planar shock wave propagating in the positive $x$-direction, see Fig. \ref{fig:wedgeGeometry}.  Given the shock Mach number, the flow properties behind the shock are uniquely given by the Rankine-Hugoniot jump conditions.

The solution is reported in non-dimensional form.  The initial density $\rho_0$ and the initial pressure $p_0$ were chosen as characteristic scales for pressure and density. The governing equations remain invariant under non-dimensionalization if the velocity is normalized by $\sqrt{p_0/\rho_0}$. In summary, the non-dimensional variables, denoted by a tilde, are: 
\begin{equation}
 \tilde{\rho}=\frac{\rho}{\rho_0} \textrm{,\ \ } 
 \tilde{p}=\frac{p}{p_0} \textrm{,\ \ }
 \tilde{u}=\frac{u}{\sqrt{\frac{p_0}{\rho_0}}} \textrm{,\ \ }
 \tilde{v}=\frac{v}{\sqrt{\frac{p_0}{\rho_0}}} .
\end{equation}

In the case of a sharp corner, the problem is self-similar and there is no characteristic length scale.   Accordingly, to report the numerical results where the grid spacing provides a characteristic scale of the numerical approximation, the $x$ and $y$ variables are such that a value of unity corresponds (arbitrarily) to one base grid spacing.  Likewise, the cases where a finite radius corner is included use the same non-dimensionalization for comparison purposes.

\subsection{Numerical method}
A numerical solution to the problem stated above was obtained by discretizing the governing equations with a finite volume scheme using Roe's linearised Riemann solver \cite{roe_1981,roe_and_pike_1984} to evaluate the fluxes.  The solution was implemented in the AMRITA computational system developed by James Quirk \cite{quirk_1998}, designed to facilitate the exchange of computational investigations among different users and adopting a philosophy where the source code is readily available for third-party scrutiny and packaged to allow verification of results obtained. The script used to run the numerical simulations is available as electronic supplementary material (Online Resource 1), and in the AMRITA newsgroup \cite{AMRITA_newsgroup}, where discussion about the computational system can also be found.
	
\begin{figure}
	\subfigure[]{
	  \includegraphics[width=\columnwidth]{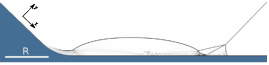}
	  \label{subfig:gridt1}
	}
	\subfigure[]{
	  \includegraphics[width=\columnwidth]{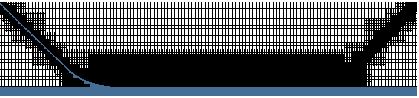}
	  \label{subfig:gridt2}
	}
\caption{Density gradient (a) and corresponding adaptive mesh (b) with four refinement levels ($M_i=6.6$, $\gamma=1.2$, ramp angle $\theta_w=44^\circ$, $R=20$)}
\label{fig:mesh}
\end{figure}

A base grid of 300 by 250 was used, aligned with the wedge wall (see Fig. \ref{fig:mesh}).  An adaptive mesh refinement technique was used to improve the accuracy of the solution near discontinuities as exemplified in Fig. \ref{fig:mesh}.  The mesh was refined around regions with density gradients, internal boundaries and behind the reflected and Mach shocks. Unless otherwise noted, four levels of mesh refinement were used, doubling the number of grid points in each direction at every level; the most refined grid spacing was thus $2^{-4}$.
  
The curved corner and upstream boundary were implemented using the embedded boundary technique described by Xu et al. \cite{xu_et_al_1997}. This permitted the use of a Cartesian computational grid.  The grid is not fitted to the curved boundary. The boundary is defined as a user-selected function.  The solution vector at the ghost cells situated across the boundary are specified to suppress flow normal to the boundary curve. 

The computations were initialised using the Rankine-Hugoniot shock solutions behind a shock discontinuity positioned eight units upstream of the start of the wedge.   However, the exact position of the shock did not influence the results.  
	
	The set of fixed parameters investigated were an isentropic exponent of $\gamma=1.2$ and incident shock Mach number of $M_i=6.6$.  These values correspond to the typical conditions of shock reflections in cellular detonation waves \cite{mach_and_radulescu_2011} with a heat release of $Q/RT_0 \approx 50$. A vast amount of literature now exists on the detonation shock reflection dynamics with these parameters (see reference \cite{mach_and_radulescu_2011} and references therein). The angle of the ramp was varied to determine the transition point between regular reflection and Mach reflection.  The
curved corner had a radius of $R=20$.
	
\subsection{Numerical verification}
\begin{figure*}
	\subfigure[$\theta_w=44^\circ$, 3 refinement levels]{
	  \includegraphics[width=0.5\textwidth]{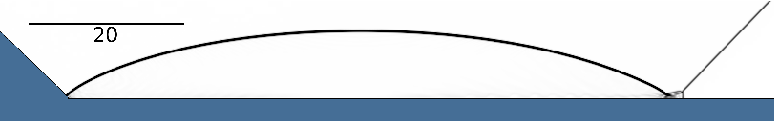}
	  \label{subfig:T44lmax3}
	}
	\subfigure[$\theta_w=45^\circ$, 3 refinement levels]{
	  \includegraphics[width=0.5\textwidth]{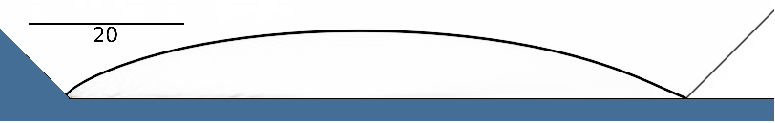}
	  \label{subfig:T45lmax3}
	}
	\subfigure[$\theta_w=44^\circ$, 4 refinement levels]{
	  \includegraphics[width=0.5\textwidth]{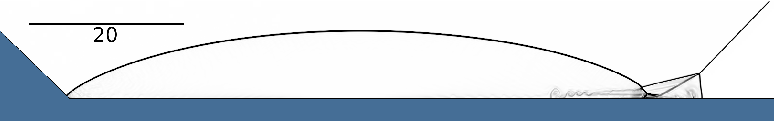}
	  \label{subfig:rampSharp44}
	}
	\subfigure[$\theta_w=45^\circ$, 4 refinement levels]{
	  \includegraphics[width=0.5\textwidth]{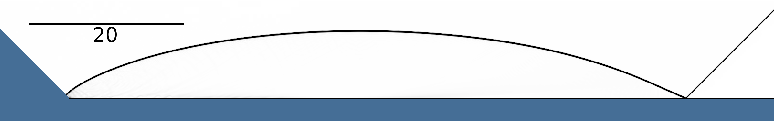}
	  \label{subfig:rampSharp45}
	}
	\subfigure[$\theta_w=44^\circ$, 5 refinement levels]{
	  \includegraphics[width=0.5\textwidth]{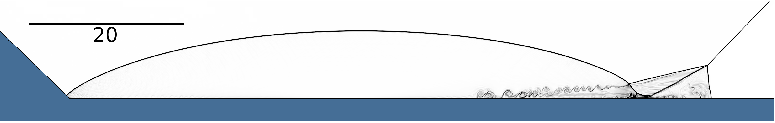}
	  \label{subfig:T44lmax5}
	}
	\subfigure[$\theta_w=45^\circ$, 5 refinement levels]{
	  \includegraphics[width=0.5\textwidth]{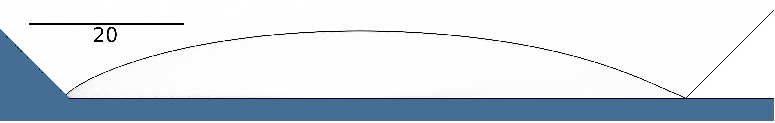}
	  \label{subfig:T45lmax5}
	}
\caption{Shock waves ($M_i=6.6$, $\gamma=1.2$) reflecting off wedges with angles at $\theta_w=44^\circ$ ((a), (c), (e), Mach reflection) and $\theta_w=45^\circ$ ((b), (d), (f), regular reflection) with sharp corners and a range of resolutions; the distance indicator refers to a distance measured in base grid spacings, as explained in the text}
\label{fig:sharpCornerResult}
\end{figure*}

To test the convergence of the results obtained with grid refinement, different resolutions were considered for the wedge with a sharp corner.  For each resolution, the wedge angle was modified in increments of $1^{\circ}$.  For a wedge angle of $\theta_w=44^\circ$, a Mach reflection was always observed with the different grid resolutions.  For a wedge angle of $\theta_w=45^\circ$, a regular reflection was always observed.  With increasing resolution, the reflection pattern remained invariant, as can be seen in Fig. \ref{fig:sharpCornerResult}. 

\begin{figure*}
\begin{center}
 \includegraphics[width=\textwidth]{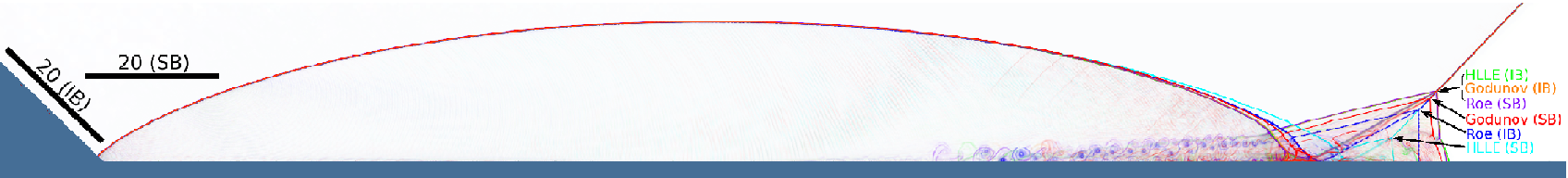}
	  \caption{Comparison of grid orientations and solvers ($M=6.6$, $\gamma=1.2$, $\theta_w=44^{\circ}$): grid aligned with wedge (SB); grid normal to incident shock / embedded-boundary wedge (IB)}
	  \label{fig:IB_vs_SB_solvers}
\end{center}
\end{figure*}

A comparison of three solvers (Roe's \cite{roe_1981,roe_and_pike_1984}, Eindfelt's Harten-Lax-van-Leer (HLLE) \cite{einfeldt_godunov-type_1991}, and Godunov's \cite{godunov_difference_1959} solvers) is shown in Fig. \ref{fig:IB_vs_SB_solvers}. The numerical experiments with the sharp-cornered wedge were also performed with different orientations of the base grid, and are superimposed on the same figure. When the base grid was aligned with the $x$-axis (normal to the incident shock), the inclined wall was captured with the embedded boundary technique.

	A range of Mach stem sizes can be seen in Fig. \ref{fig:IB_vs_SB_solvers}, depending on the orientation and solver used. Nevertheless, the differences between the results obtained with both grid orientations were relatively small for a wedge angle near the transition angle. The solution obtained with both grids show the same flow features.  These errors are potentially due to approximations of the embedded boundary technique.  For this reason, and to minimize such errors, the grid was oriented with the wedge wall.	The Godunov and HLLE solvers used in conjunction with the embedded boundary wedge gave very similar results to the Roe solver used with the grid-oriented wedge. The results shown in this paper were achieved using Roe's solver with the grid-aligned wedge.

%---------------------------------------------------------------------------------------------------------------------
\section{Results}
\label{sec:results}
\begin{figure}
\begin{center}
 \includegraphics[width=0.7\columnwidth]{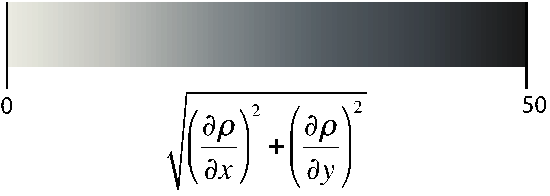}
	  \caption{Grayscale for rendering the numerical results of density gradients.}
	  \label{fig:grayscale}
\end{center}
\end{figure}

The results of the numerical calculations are presented below as pseudo-schlieren records.  The absolute value of the density gradients is shown according to the grey-scale given in Fig. \ref{fig:grayscale}.

%---------------------------------------------------------------------------------------------------------------------
\subsection{Reflection over a wedge with a sharp corner}
\label{sec:sharpCorner}
Figure \ref{fig:sharpCornerResult} illustrates the resulting flow-fields obtained for a shock reflecting off a ramp with an angle of $\theta_w=44^\circ$ and $\theta_w=45^\circ$ and a sharp corner.  The domain has been cropped to show the area of interest.  As discussed above, at $\theta_w=45^\circ$, a regular reflection is observed, while at $\theta_w=44^\circ$, the reflection takes the form of a Mach reflection, characterised by a Mach shock travelling nearly perpendicular to the wall. A second triple-shock configuration (triple-point), also with an associated slip line, which becomes unstable due to the Kelvin-Helmholtz instability, can be found along the reflected wave and corresponds to the sub-regime of double Mach reflection \cite{hornung_regular_1986}. This double Mach reflection has a negative reflection angle (between the triple-point path and reflected wave) \cite{gvozdeva_et_al_2012}.

%---------------------------------------------------------------------------------------------------------------------
\subsection{Reflection over a wedge with a rounded corner}
\label{sec:roundCorner}
\begin{figure*}
	\subfigure[$\theta_w=44^\circ$]{
	  \includegraphics[width=\textwidth]{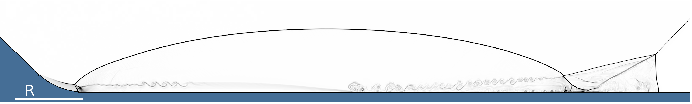}
	  \label{subfig:AR44r20}
	}
	\subfigure[$\theta_w=51^\circ$]{
	  \includegraphics[width=\textwidth]{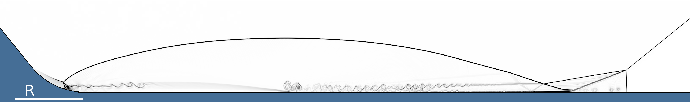}
	  \label{subfig:AR50r20}
	}
	\subfigure[$\theta_w=57^\circ$]{
	  \includegraphics[width=\textwidth]{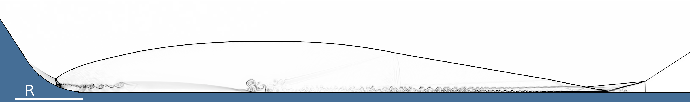}
	  \label{subfig:AR55r20}
	}
	\subfigure[$\theta_w=58^\circ$]{
	  \includegraphics[width=\textwidth]{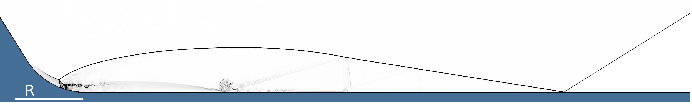}
	  \label{subfig:AR60r20}
	}
\caption{Shock waves ($M_i=6.6$, $\gamma=1.2$, $R=20$) reflecting off wedges with angle (a) $\theta_w=44^\circ$, (b) $\theta_w=51^\circ$, (c) $\theta_w=57^\circ$, (d) $\theta_w=58^\circ$, and curved corners}
\label{fig:roundCornerResult}
\end{figure*}

Figure \ref{fig:roundCornerResult} shows the flow-fields of the shock reflection over a wedge with a rounded corner after the reflection point has travelled a distance approximately eight times the radius of the corner.  Double Mach reflections are seen at ramp angles of $\theta_w=44^\circ$, $\theta_w=51^\circ$, and $\theta_w=57^\circ$. A regular reflection appears at a ramp angle of $\theta_w=58^\circ$.  The transition from Mach to regular reflection now occurs at a significantly larger angle ($\theta_w=57.5^\circ \pm 0.5^\circ$) as compared with the critical wedge angle for transition with a sharp corner ($\theta_w=44.5^\circ \pm 0.5^\circ$). 

\begin{figure*}
	\subfigure[$\theta_w=57^\circ$]{
	  \includegraphics[width=0.5\textwidth]{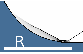}
	  \label{subfig:T57-1}
	}
	\subfigure[$\theta_w=58^\circ$]{
	  \includegraphics[width=0.5\textwidth]{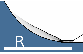}
	  \label{subfig:T58-1}
	}	
	\subfigure[$\theta_w=57^\circ$]{
	  \includegraphics[width=0.5\textwidth]{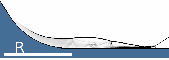}
	  \label{subfig:T57-2}
	}
	\subfigure[$\theta_w=58^\circ$]{
	  \includegraphics[width=0.5\textwidth]{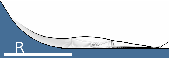}
	  \label{subfig:T58-2}
	}
	\subfigure[$\theta_w=57^\circ$]{
	  \includegraphics[width=0.5\textwidth]{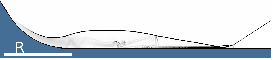}
	  \label{subfig:T57-3}
	}
	\subfigure[$\theta_w=58^\circ$]{
	  \includegraphics[width=0.5\textwidth]{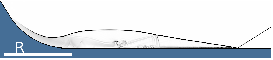}
	  \label{subfig:T58-3}
	}
	\subfigure[$\theta_w=57^\circ$]{
	  \includegraphics[width=0.5\textwidth]{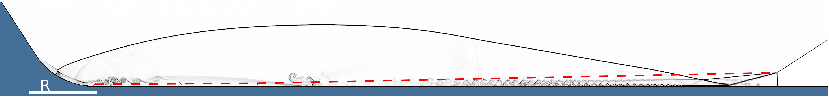}
	  \label{subfig:T57-4}
	} 
	\subfigure[$\theta_w=58^\circ$]{
	  \includegraphics[width=0.5\textwidth]{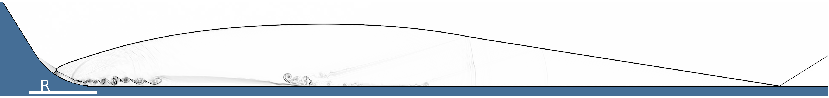}
	  \label{subfig:T58-4}
	}   
\caption{Evolution of a shock wave ($M_i=6.6$, $\gamma=1.2$, $R=20$) reflecting off a $\theta_w=57^\circ$ wedge ((a), (c), (e), (g)) and $\theta_w=58^\circ$ wedge ((b), (d), (f), (h)) with rounded corners; triple-point path has been superimposed (dashed-line) on sub-figure (g)}
\label{fig:roundCornerEvolution}
\end{figure*}

\begin{figure}
	  \includegraphics[width=0.50\textwidth]{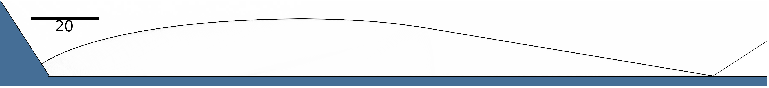} 
\caption{Shock wave ($M_i=6.6$, $\gamma=1.2$) reflecting off a wedge $\theta_w=57^\circ$ with a sharp corner}
\label{fig:T50sharp}
\end{figure}

	The evolutions of reflections for wedge angles of $\theta_w=57^\circ$ and $\theta_w=58^\circ$ with curved corners are shown in Fig. \ref{fig:roundCornerEvolution}. For the wedge angle $\theta_w=57^\circ$, the reflection process asymptotes to a new self-similar solution, a (double) Mach reflection, as the shock travels a large distance from the rounded corner. This differs from the one obtained over a sharp-cornered wedge at the same angle shown in Fig. \ref{fig:T50sharp}, which was a regular reflection. The reflection over the $\theta_w=58^\circ$ wedge keeps the form of a Mach reflection for approximately two-and-a-half corner radii after tangency with the wedge. The triple point then collides with the wedge to become a regular reflection.  The trajectory of the triple-point for $\theta_w=57^\circ$ is shown in Fig. \ref{subfig:T57-4}.   The triple point moves away from the wedge and does not appear to collide with it at a distance of up to ten times that of the radius of curvature. It is evident that the new asymptotically self-similar solution consists of a Mach reflection which grows with time in a linear fashion.   

Although the results reported were only for a distinct set of parameters in gases with high shock strength and low isentropic exponent typical of gaseous detonations, similar results and conclusions were also obtained for weaker shock reflections with an isentropic exponent corresponding to a diatomic gas like air.  Numerical experiments with $\gamma=1.4$ and $M_i=3$, for example, revealed a transition angle from Mach to regular reflection at $\theta_w=50.5^\circ \pm 0.5^\circ $.  With an initially curved wedge, Mach reflections were maintained for larger wedge angles.

%---------------------------------------------------------------------------------------------------------------------
\section{Discussion}
\label{sec:discussion}

For parameters $M_i=6.6$ and $\gamma=1.2$, the determined transition between regular and Mach reflection of $44.5^\circ\pm0.5^\circ$ over the wedge with a sharp corner is in good agreement with the sonic criterion \cite{hornung_et_al_1979}, modeled by the two-shock theory.  This criterion  predicts a transition at $\theta_w=44.47^\circ \pm 0.01^\circ$, while the detachment criterion predicts a transition at $\theta_w=44.42^\circ \pm 0.01^\circ$.  The transition from Mach to regular reflection occurs when the flow behind the reflected shock becomes outside the range of influence of the corner (i.e., when the flow behind the reflection point is supersonic. Thus, it appears that the self-similar solution is indeed established such that the shock reflection retains its shape as it grows with time. 

However, the numerical experiments conducted with the curved corner illustrate that the same self-similar solution is not recovered in the far field, when the corner is initially curved.  A possible explanation for this is likely also related to the sonic criterion, see below.  In a progressively curved ramp, the reflection starts out invariably as a Mach reflection due to the initial small angle.  As the wedge angle increases continuously along the curved corner, the Mach stem likely remains strong enough to communicate a scale of length; the entire reflection configuration is subject to the changing angle \cite{krasovskaya_and_berezkina_2008}. The  shock reflection is then likely able to maintain this subsonic pathway past the corner along the flat portion of the wedge.  Future studies could address this mechanism by tracking the path of acoustic signals. 

It is also instructive to consider the present results in view of previous studies performed with modifications to the straight wedge problem.  The problem of unsteady shock reflection over curved surfaces has been addressed by a number of authors \cite{henderson_and_lozzi_1979,ben-dor_and_takayama_1985,itoh_et_al_1981,ben-dor_et_al_1980}.  These studies did not address the continuation of the curved surface into a straight surface; i.e., the asymptotically self-similar regime studied here.  Instead, they were concerned with the transition of the Mach reflection to regular reflection over the curved wedge, which occurs when the triple-point trajectory eventually merges with the curved wall.  Similar to the present study, it was found that the wedge angle at which the transition from Mach to regular reflection occurs is larger than the angle obtained for reflection on a straight wedge. It has also been shown that the transition angle decreases with radius of curvature \cite{takayama_and_sasaki_1983}, tending towards the transition angle of a reflection off a straight wedge. The transition angle for a reflection over a concave circle has been analytically predicted using the sonic criterion \cite{hornung_et_al_1979} by Ben-Dor and Takayama \cite{ben-dor_and_takayama_1985} but the proposed models are independent of the radius of curvature. A model incorporating the radius of curvature remains yet incomplete \cite{takayama_and_ben-dor_1986}. Their model ``A" predicts a transition at $\theta_w \approx 63^\circ$ for $M=6.6$ and $\gamma=1.2$ on the concave circle.  The value derived from Ben-Dor and Takayama's model is higher than what was obtained for a curved wall followed by a straight wedge in this study, namely $\theta_w=57.5^\circ \pm 0.5$.  This difference is an expected result due to slight difference between the two problems.  For the initially curved wedge in this study, the Mach reflection vanishes when the trajectory of the triple point intersects the straight part of the wedge, as shown in Fig. \ref{fig:roundCornerEvolution}.  In the continuously curved wall, the Mach reflection vanishes when the trajectory of the triple intersects the curved part of the wedge, which, from simple geometric arguments occurs for larger angles than in the problem studied here.  Nevertheless, the similarity in results suggests that the persistence of the subsonic pathway for the curved wall, used by Ben-Dor and Takayama to model their results, is the likely explanation for the delayed transition to regular reflection.  

Another problem analogous to the present is the shock reflection over two sequential straight wedges, where the angle of the second wedge is greater than the first.  This problem has been studied by Ben-Dor et al. \cite{ben-dor_et_al_1987,ben-dor_et_al_1988}.  However, it was found that the reflection pattern quickly recovered the configuration expected had the first wedge been absent.  	

%---------------------------------------------------------------------------------------------------------------------
\section{Conclusion}
\label{sec:conc}
The present study shows unambiguously that the details of the start-up of an asymptotically self-similar solution play a very important role in determining the final asymptotic self-similar solution.  The shock reflection studied showed that the solution achieved from reflection past a straight wedge, which starts with a finite curvature, persists as a Mach reflection in the far-field through wedge angles far superior to those observed over sharp-cornered wedges.	As the radius of the rounded corner becomes insignificant in comparison to the distance travelled by the incident shock, the solution of the asymptotically self-similar problem disagrees with the solution of the self-similar problem. 

The curved wedge corner introduces a length scale to the problem in a similar way that viscous effects or shock relaxation phenomena would. While viscous effects \cite{hornung_and_taylor_1982,ben-dor_1987} and the presence of relaxation reduce the transition wedge angle \cite{hornung_et_al_1979}, these length scales vanish asymptotically as the reflection grows. Their effects only dominate the solution at early times, while the later time solution may still be influenced by the start-up conditions, as illustrated in the present study.  Indeed, the present study shows that the flow patterns obtained from the start-up conditions persist through to scales significantly larger than those at the corner, and play a dominant role determining what is obtained into the asymptotic pseudosteady configuration. Similar results have been reported by Henderson and Menikoff \cite{henderson_and_menikoff_1998} with regards to shock reflections in steady-state flows.

In view of the present findings, it appears worthwhile to re-evaluate other self-similar problems of gas dynamics, such as Taylor-Sedov blast waves, or the problem of shock diffraction at a sudden area change, to determine the influence of the starting conditions on the solution obtained in the far field.  This issue is particularly important in view of current numerical simulation techniques used to solve these problems, which attempt to regularize the start-up conditions to resolve them with a finite size computational grid.  

%---------------------------------------------------------------------------------------------------------------------
\begin{acknowledgements}
The authors wish to thank James Quirk for his support with the AMRITA system. 

This work was sponsored by a Natural Sciences and Engineering Research Council of Canada (NSERC) Discovery grant to M. I. Radulescu, with further support from the NSERC H2CAN Strategic Network of Excellence, and a NSERC Alexander Graham Bell Canadian Graduate Scholarship to S. S.M. Lau-Chapdelaine.
\end{acknowledgements}


\begin{thebibliography}{}
% etc


\bibitem{ben-dor_shock_2007}
	Ben-Dor, G.: Shock wave reflection phenomena. Springer, (2007)
	
\bibitem{semenov_et_al_2009a}
	Semenov, A. N., Berezkina, M. K., Krasovskaya, I. V.: Classification of shock wave reflections from a wedge. Part 1 : Boundaries and domains of existence for different types of reflections. Tech. Phys., 54, 4, 491-496 (2009)

\bibitem{semenov_et_al_2009b}	
	Semenov, A. N., Berezkina, M. K., Krasovskaya, I. V.: Classification of shock wave reflections from a wedge. Part 2 : Experimental and numerical simulations of different types of Mach reflections. Tech. Phys., 54, 4, 497-503 (2009)

\bibitem{vasilev_et_al_2008}
	Vasilev, E. I., Elperin, T., Ben-Dor, G.: Analytical reconsideration of the von Neumann paradox in the reflection of a shock wave over a wedge. Phys. Fluids, 20, 046101 (2008)

\bibitem{hornung_regular_1986}
	Hornung, H.: Regular and Mach reflection of shock waves. Annu. Rev. Fluid Mech., 18, 33-58 (1986)

\bibitem{henderson_and_menikoff_1998}
	Henderson, L. F., Menikoff, R.: Triple-shock entropy theorem and its consequences. J. Fluid Mech., 366, 179-210 (1998)
	
\bibitem{henderson_and_lozzi_1975}
	Henderson, L. F., Lozzi, A.: Experiments on transition of Mach reflexion. J. Fluid Mech., 68, 139-155 (1975)

\bibitem{henderson_and_lozzi_1979}
	Henderson, L. F., Lozzi, A.: Further experiments on transition to Mach reflexion. J. Fluid Mech., 94, 541-559 (1979)
	
\bibitem{hornung_et_al_1979}
	Hornung, H. G., Oertel, H., Sandeman, R. J.: Transition to Mach reflexion of shock waves in steady and pseudosteady flow with and without relaxation. J. Fluid Mech., 90, 541-560 (1979)

\bibitem{jones_et_al_1951}
	Jones, D. M., Martin P., Thornhill C. K.: A note on the pseudo-stationary flow behind a strong shock diffracted or reflected at a corner. Proc. Roy. Soc. Lond. A, 209, 238-248 (1951)

\bibitem{henderson_2001}
	Henderson, L. F.: Handbook of Shock Waves, vol. 2, Chapter 8.2, Two-Dimensional Interactions: The Refraction of Shock Waves. Academic Press, 181-203 (2001)

\bibitem{landau_and_lifshitz_1987}
	Landau, L. D., Lifshitz, E. M.: Course of Theoretical Physics, Volume 6: Fluid Mech., Second English Edition. Pergamon Press, 403-411 (1987)

\bibitem{ben-dor_and_takayama_1985}
	Ben-Dor, G., Takayama, K.: Analytical prediction of the transition from Mach to regular reflection over cylindrical concave wedges. J. Fluid Mech., 158, 365-380 (1985)

\bibitem{itoh_et_al_1981}
	Itoh, S. Okazaki, N., Itaya, M.: On the transition between regular and Mach reflection in truly non-stationary flows. J. Fluid Mech., 108, 383-400 (1981)

\bibitem{ben-dor_et_al_1980}
	Ben-Dor, G., Takayama, K., Kawauchi, T.: The transition from regular to Mach reflexion and from Mach to regular reflexion in truly non-stationary Flows. J. Fluid Mech., 100, 147-160 (1980)

\bibitem{krasovskaya_and_berezkina_2008}
	Krasovskaya, I. V., Berezkina, M. K.: On reflection of shock waves and shock-wave configurations. Tech. Phys. Lett., 34, 177-179, (2008)

\bibitem{roe_1981}
	Roe, P. L.: Approximate Riemann solvers, parameter vectors, and difference schemes. J. Comput. Phys., 43, 357-372 (1981)

\bibitem{roe_and_pike_1984}
	Roe, P. L., Pike, J.: Efficient construction and utilisation of approximate Riemann solutions. In: Computer methods in application science and engineering VI. Proceedings 6th International Symposium, Versailles, France; Netherlands, pp. 499-518 (1984)
	
\bibitem{quirk_1998}
	Quirk, J. J.: AMRITA. VKI 29th CFD lecture series, Febuary 1998

\bibitem{AMRITA_newsgroup}
	Quirk, J. J.: AMRITA ebook $>$ shock/detonation interaction with a ramp. Google Groups. https://groups.google.com/d/msg/amrita-ebook/tKmkJyL4TC0/xn9vrnFcG-EJ (2013), available through http://amrita-ebook.org/. Accessed 25 April 2013

\bibitem{xu_et_al_1997}
	Xu, S., Aslam, T., Stewart, D. S.: High resolution numerical simulation of ideal and non-ideal compressible reacting flows with embedded internal boundaries. Combust. Theory Model., 1, 113-142 (1997)
	
\bibitem{mach_and_radulescu_2011}
	Mach, P., Radulescu, M. I.: Mach reflection bifurcations as a mechanism of cell multiplication in gaseous detonations. Proc. Combust. Inst., 33, 2279-2285 (2011)

\bibitem{einfeldt_godunov-type_1991}
	Einfeldt, B., Munz, C. D., Roe, P. L., Sj\"ogreen, B.: On Godunov-type methods near low densities. J. Comput. Phys., 92, 273-295 (1991)

\bibitem{godunov_difference_1959}
	Godunov, S.K.: A difference method for numerical calculation of discontinuous solutions of the equations of hydrodynamics. Matematicheskii Sbornik, 89, 271-306 (1959)

\bibitem{gvozdeva_et_al_2012}
	Gvozdeva, G. L., Borsch, V. L., Gavrenkov, S. A.: Analytical and Numerical Study of Three Shock Configurations with Negative Reflection Angle. In: Kontis, K. (ed.), 28th International Symposium Shock Waves, pp. 587-592. Springer, Berlin Heidelberg (2012)

\bibitem{takayama_and_sasaki_1983}
	Takayama, K., Sasaki, M.: Effects of radius of curvature and initial angle on the shock transition over concave or convex walls: Memorial Institute of High Speed Mechanics, Tohoku University., 46: 1-30 (1983)

\bibitem{takayama_and_ben-dor_1986}
	Takayama, K., Ben-Dor, G.: Reflection and diffraction of shock waves over circular concave wall. Reports of the  Institute of High Speed Mechanics, Tohoku University, 51: 43-87 (1986)

\bibitem{ben-dor_et_al_1987}
	Ben-Dor, G., Dewey, J. M., Takayama K.: The reflection of a plane shock over a double wedge: J. Fluid Mech., 176, 483-520 (1987)

\bibitem{ben-dor_et_al_1988}
	Ben-Dor, G., Dewey, J. M., McMillin, D. J., Takayama, K.: Experimental investigation of the asymptotically approached Mach reflection over the second surface in a double wedge reflection: Exp. Fluids, 6, 429-434 (1988)

\bibitem{hornung_and_taylor_1982}
	Hornung, H. G., Taylor, J. R.: Transition from regular to Mach reflection of shock waves Part 1. The effect of viscosity in the pseudosteady case. J. Fluid Mech., 123, 143-153 (1982)
	
\bibitem{ben-dor_1987}
	Ben-Dor, G.: A reconsideration of the three-shock theory for a pseudo-steady Mach reflection. J. Fluid Mech., 181, 467-484 (1987)

\end{thebibliography}


\end{document}